\documentstyle[psfrag,preprint,graphicx,tighten,eqsecnum,floats,aps]{revtex}

\def\l{\ell_{\rm Pl}}
\def\P{{\cal P}}
\def\N{{\cal N}}

\def\H{{\cal H}}
\def\Hb{{\cal H}_{\rm bulk}}
\def\Hs{{\cal H}_{\rm surface}}
\def\scri{{\cal I}}
\def\spi{i^o}

\preprint{\vbox{\baselineskip=12pt
\rightline{CGPG-99/8-4}}}

\begin{document}
\draft
\title{Interface of General Relativity, Quantum Physics and\\
Statistical Mechanics: Some Recent Developments} 
\author {Abhay\ Ashtekar}
\address{Center for Gravitational Physics and Geometry \\
Department of Physics, The Pennsylvania State University \\
104 Davey, University Park, PA 16802, USA}

\maketitle

\begin{abstract}

The arena normally used in black holes thermodynamics was recently
generalized to incorporate a broad class of physically interesting
situations.  The key idea is to replace the notion of stationary event
horizons by that of `isolated horizons.'  Unlike event horizons,
isolated horizons can be located in a space-time
\textit{quasi-locally}.  Furthermore, they need not be Killing
horizons. In particular, a space-time representing a black hole which
is itself in equilibrium, but whose exterior contains radiation,
admits an isolated horizon.  In spite of this generality, the zeroth
and first laws of black hole mechanics extend to isolated horizons.
Furthermore, by carrying out a systematic, non-perturbative
quantization, one can explore the quantum geometry of isolated
horizons and account for their entropy from statistical mechanical
considerations. After a general introduction to black hole
thermodynamics as a whole, these recent developments are briefly
summarized.
\end{abstract}

\section{Motivation}
\label{sec1}

In the seventies, there was a flurry of activity in black hole physics
which brought out an unexpected interplay between general relativity,
quantum field theory and statistical mechanics \cite{1,2,3,4}.  That
analysis was carried out only in the semi-classical approximation,
i.e., either in the framework of Lorentzian quantum field theories in
curved space-times or by keeping just the leading order, zero-loop
terms in Euclidean quantum gravity.  Nonetheless, since it brought
together the three pillars of fundamental physics, it is widely
believed that these results capture an essential aspect of the more
fundamental description of Nature.  For over twenty years, a concrete
challenge to all candidate quantum theories of gravity has been to
derive these results from first principles, without invoking
semi-classical approximations.

Specifically, the early work is based on a somewhat ad-hoc mixture of
classical and semi-classical ideas ---reminiscent of the Bohr model of
the atom--- and generally ignored the quantum nature of the
gravitational field itself.  For example, statistical mechanical
parameters were associated with macroscopic black holes as follows.
The laws of black hole mechanics were first derived in the framework
of {\it classical} general relativity, without any reference to the
Planck's constant $\hbar$ \cite{2}.  It was then noted that they
have a remarkable similarity with the laws of thermodynamics if one
identifies a multiple of the surface gravity $\kappa$ of the black
hole with temperature and a corresponding multiple of the area $a_{\rm
hor}$ of its horizon with entropy.  However, simple dimensional
considerations and thought experiments showed that the multiples must
involve $\hbar$, making quantum considerations indispensable for a
fundamental understanding of the relation between black hole mechanics
and thermodynamics \cite{1}.  Subsequently, Hawking's investigation of
(test) quantum fields propagating on a black hole geometry showed that
black holes emit thermal radiation at temperature $T_{\rm rad} =
\hbar\kappa/2\pi$ \cite{3}.  It therefore seemed natural to assume
that black holes themselves are hot and their temperature $T_{\rm bh}$
is the same as $T_{\rm rad}$.  The similarity between the two sets of
laws then naturally suggested that one associate an entropy $S_{\rm bh}
= a_{\rm hor}/4\hbar$ with a black hole of area $a_{\rm hor}$.  While
this procedure seems very reasonable, it does not provide a
`fundamental derivation' of the thermodynamic parameters $T_{\rm bh}$
and $S_{\rm bh}$.  The challenge is to derive these formulas from
first principles, i.e., by regarding large black holes as statistical
mechanical systems in a suitable quantum gravity framework.

Recall the situation in familiar statistical mechanical systems such
as a gas, a magnet or a black body.  To calculate their thermodynamic
parameters such as entropy, one has to first identify the elementary
building blocks that constitute the system.  For a gas, these are
molecules; for a magnet, elementary spins; for the radiation field in
a black body, photons.  What are the analogous building blocks for
black holes?  They can not be gravitons because the underlying
space-times were assumed to be stationary.  Therefore, the elementary
constituents must be non-perturbative in the terminology of local
field theory.  Thus, to account for entropy from first principles
within a candidate quantum gravity theory, one would have to: i)
isolate these constituents; ii) show that, for large black holes, the
number of quantum states of these constituents goes as the exponential
of the area of the event horizon; and, iii) account for the Hawking
radiation in terms of processes involving these constituents and
matter quanta.

These are difficult tasks, particularly because the very first step
--isolating the relevant constituents-- requires new conceptual as
well as mathematical inputs.  Furthermore, in the semi-classical
theory, thermodynamic properties have been associated not only with
black holes but also with cosmological horizons.  Therefore, ideally,
the framework has to be sufficiently general to encompass these
diverse situations.  It is only recently, more than twenty years after
the initial flurry of activity, that detailed proposals have emerged.
The more well-known of these comes from string theory \cite{ms} where
the relevant elementary constituents are associated with D-branes
which lie outside the original perturbative sector of the theory.  The
purpose of this contribution is to summarize the ideas and results
from another approach which emphasizes the quantum nature of geometry,
using non-perturbative techniques from the very beginning.  Here, the
elementary constituents are the quantum excitations of geometry itself
and the Hawking process now corresponds to the conversion of the
quanta of geometry to quanta of matter.  Although the two approaches
seem to be strikingly different from one another, as I will indicate,
in a certain sense they are complementary.

\section{Key Issues}
\label{s2}
 
In the last section, I focussed on quantum issues. However, the status
of \textit{classical} black hole mechanics, which provided much of the
inspiration in quantum considerations, has itself remained
unsatisfactory in some ways.  Therefore, in a systematic approach, one
has to revisit the classical theory before embarking on quantization.

The zeroth and first laws of black hole mechanics refer to equilibrium
situations and small departures therefrom.  Therefore, in this
context, it is natural to focus on isolated black holes.  However, in
standard treatments, these are generally represented by
\textit{stationary} solutions of field equations, i.e, solutions which
admit a time-translation Killing vector field \textit{everywhere},
not just in a small neighborhood of the black hole.  While this simple
idealization is a natural starting point, it seems to be overly
restrictive.  Physically, it should be sufficient to impose boundary
conditions at the horizon which ensure \textit{only the black hole
itself is isolated}.  That is, it should suffice to demand only that
the intrinsic geometry of the horizon be time independent, whereas the
geometry outside may be dynamical and admit gravitational and other
radiation.  Indeed, we adopt a similar viewpoint in ordinary
thermodynamics; in the standard description of equilibrium
configurations of systems such as a classical gas, one usually assumes
that only the system under consideration is in equilibrium and
stationary, not the whole world.  For black holes, in realistic
situations one is typically interested in the final stages of
collapse where the black hole is formed and has `settled down' or in
situations in which an already formed black hole is isolated for the
duration of the experiment (see figure 1).  In such situations, there
is likely to be gravitational radiation and non-stationary matter far
away from the black hole, whence the space-time as a whole is not
expected to be stationary.  Surely, black hole mechanics should
incorporate in such situations.

\begin{figure}
  \begin{center}
    \begin{minipage}{2.5in}
      \begin{center}
        \psfrag{Ip}{$\scri^+$}
    	\psfrag{calM}{$\mathcal{M}$}
    	\psfrag{ip}{$i^+$}
	\psfrag{i0}{$\spi$}
        \psfrag{Delta}{$\Delta$}
        \psfrag{M}{$M$}
        \includegraphics[height=6cm]{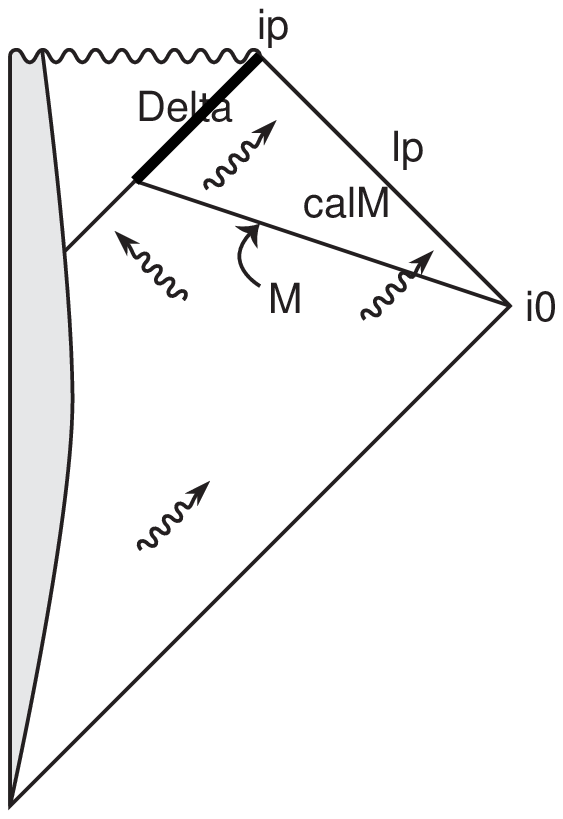}\\(a)
      \end{center}
    \end{minipage}
    \hspace{.5in}
    \begin{minipage}{3in}
      \begin{center}
        \includegraphics[width=6cm]{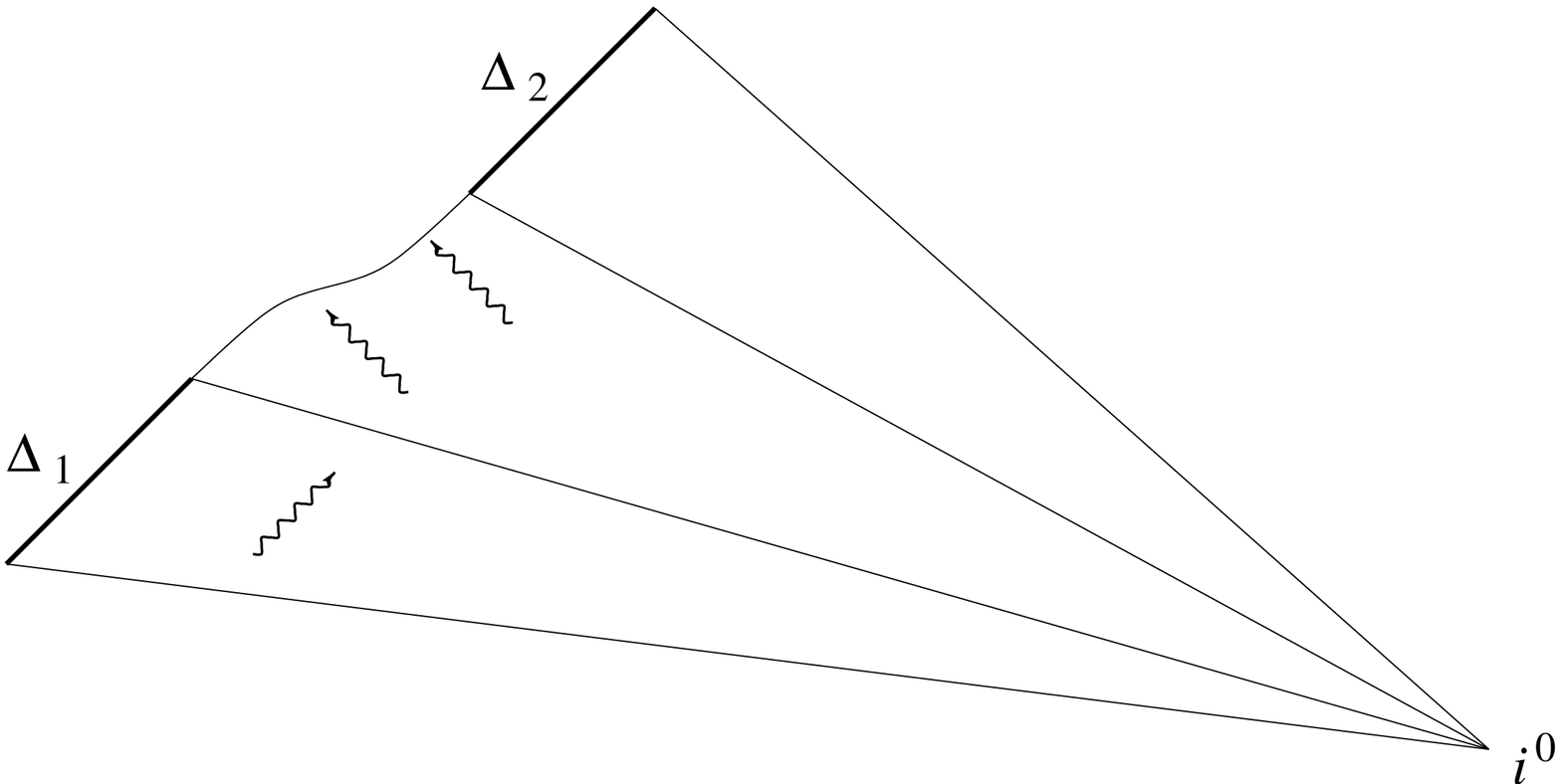}\\(b)
      \end{center}
    \end{minipage}
\caption{(a)\quad A typical gravitational collapse.
The portion $\Delta$ of the horizon at late times is isolated.  The
space-time $\mathcal{M}$ of interest is the triangular region bounded
by $\Delta$, $\scri^+$ and a partial Cauchy slice $M$.  \quad(b)\quad
Space-time diagram of a black hole which is initially in equilibrium,
absorbs a small amount of radiation, and again settles down to
equilibrium.  Portions $\Delta_1$ and $\Delta_2$ of the horizon are
isolated.}\label{exam}
\end{center}
\end{figure}

A second limitation of the standard framework lies in its dependence
on \textit{event} horizons which can only be constructed
retroactively, after knowing the \textit{complete} evolution of
space-time.  Consider for example, Figure 2 in which a spherical star
of mass $M$ undergoes a gravitational collapse.  The singularity is
hidden inside the null surface $\Delta_1$ at $r=2M$ which is foliated
by a family of marginally trapped surfaces and would be a part of the
event horizon if nothing further happens.  Suppose instead, after a
very long time, a thin spherical shell of mass $\delta M$ collapses.
Then $\Delta_1$ would not be a part of the event horizon which would
actually lie slightly outside $\Delta_1$ and coincide with the surface
$r= 2(M+\delta M)$ in distant future.  On physical grounds, it seems
unreasonable to exclude $\Delta_1$ a priori from thermodynamical
considerations.  Surely one should be able to establish the standard
laws of laws of mechanics not only for the event horizon but also for
$\Delta_1$.

Another example is provided by cosmological horizons in de Sitter
space-time \cite{4}.  In this case, there are no singularities or
black-hole event horizons.  On the other hand, semi-classical
considerations enable one to assign entropy and temperature to these
horizons as well.  This suggests the notion of event horizons is too
restrictive for thermodynamical analogies.  We will see that this is
indeed the case; as far as equilibrium properties are concerned, the
notion of event horizons can be replaced by a more general,
quasi-local notion of `isolated horizons' for which the familiar laws
continue to hold.  The surface $\Delta_1$ in figure 2 as well as the
cosmological horizons in de Sitter space-times are examples of
isolated horizons.

\begin{figure}
  \begin{center}
    \includegraphics[height=4cm]{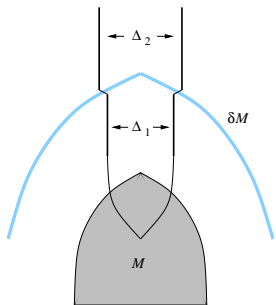}
\caption{A spherical star of mass $M$ undergoes collapse.  Later, a
spherical shell of mass $\delta{M}$ falls into the resulting black
hole.  While $\Delta_1$ and $\Delta_2$ are both isolated horizons,
only $\Delta_2$ is part of the event horizon.}\label{shell}
\end{center}
\end{figure}

At first sight, it may appear that only a small extension of the
standard framework, based on stationary event horizons, is needed to
overcome the limitations discussed above.  However, this is not the
case.  For example, in the stationary context, one identifies the
black-hole mass with the ADM mass defined at spatial infinity.  In the
presence of radiation, this simple strategy is no longer viable since
radiation fields well outside the horizon also contribute to the ADM
mass.  Hence, to formulate the first law, a new definition of the
black hole mass is needed.  Similarly, in the absence of a global
Killing field, the notion of surface gravity has to be extended in a
non-trivial fashion.  Indeed, even if space-time happens to be static
in a neighborhood of the horizon ---already a stronger condition than
contemplated above--- the notion of surface gravity is ambiguous
because the standard expression fails to be invariant under constant
rescalings of the Killing field.  When a \textit{global} Killing field
exists, the ambiguity is removed by requiring the Killing field be
unit at \textit{infinity}.  Thus, contrary to intuitive expectation,
the standard notion of the surface gravity of a stationary black hole
refers not just to the structure at the horizon, but also to infinity.
This `normalization problem' in the definition of the surface gravity
seems especially difficult in the case of cosmological horizons in
(Lorentzian) space-times whose Cauchy surfaces are compact.  Apart
from these conceptual problems, a host of technical issues must also
be resolved.  In Einstein-Maxwell theory, the space of stationary
black hole solutions is three dimensional whereas the space of
solutions admitting isolated horizons is \textit{infinite}-dimensional
since these solutions admit radiation near infinity.  As a result, new
techniques have to be used and these involve some functional analytic
subtleties.

This set of issues has a direct bearing on quantization as well. For,
in a systematic approach, one would first extract an appropriate
sector of the theory in which space-time geometries satisfy suitable
conditions at interior boundaries representing horizons, then
introduce a well-defined action principle tailored to these boundary
conditions, and, finally, use the resulting Lagrangian or Hamiltonian
frameworks as points of departure for constructing the quantum
theory. If one insists on using \textit{event} horizons, these steps
are difficult to carry out because the resulting boundary conditions
do not translate in to (quasi-)local restrictions on fields. Indeed,
for event horizon boundaries, there is \textit{no} action principle
available in the literature.  The restriction to \textit{globally}
stationary space-times causes additional difficulties. For, by no hair
theorems, the space of stationary solutions admitting event horizons
is finite dimensional and quantization of this `mini-superspace' would
ignore all field theoretic effects \textit{by fiat}. Indeed, most
treatments of black hole mechanics are based on differential geometric
identities and field equations, and are not at all concerned with such
issues related to quantization.

Thus, the first challenge is to find a new framework which achieves,
in a single stroke, three goals: i) it overcomes the two limitations
of black hole mechanics by finding a better substitute for stationary
event horizons; ii) generalizes laws of black hole mechanics to the
new, more physical paradigm; and, iii) leads to a well-defined action
principle and Hamiltonian framework which can serve as spring-boards
for quantization. The second challenge is then to: i) carry out
quantization non-perturbatively; ii) obtain a quantum description of
the horizon geometry; and, iii) account for the the horizon entropy
statistical mechanically by counting the underlying micro-states. As
discussed in the next section, these goals have been met for
non-rotating isolated horizons.

\section{Summary}
\label{s3}

In this section, I will sketch the main ideas and results on the
classical and quantum physics of isolated horizons and provide
a guide to the literature where details can be found.

\subsection{Isolated horizons}
\label{s3.1}

The detailed boundary conditions defining non-rotating isolated
horizons were introduced in \cite{ack,abf2}. Basically, an isolated
horizon $\Delta$ is a null 3-surface, topologically $S^2\times R$,
foliated by a family of marginally trapped 2-spheres. Denote the
normal direction field to $\Delta$ by $[\ell^a]$. Being null, it is
also tangential to $\Delta$. The boundary conditions require that it
be expansion-free, so that the area of the marginally trapped surface
remains constant `in time'. Assuming that the matter fields under
consideration satisfy a very weak `energy condition' at $\Delta$, the
Raychaudhuri equation then implies that there is no flux of matter
across $\Delta$. More detailed analysis also shows that there is no
flux of gravitational radiation. (More precisely, the Newman-Penrose
curvature component $\Psi_0$ vanishes on $\Delta$.) These properties
capture the idea that the horizon is isolated. Denote the second null
normal to the family of marginally trapped 2-spheres by $[n^a]$. There
are additional conditions on the Newman-Penrose spin coefficients
associated with $[n^a]$ which ensure that $\Delta$ is a
\textit{future} horizon with no rotation.

Event horizons of static black holes of the Einstein-Maxwell-Dilaton
theory are particular examples of non-rotating isolated horizons. The
cosmological horizons in de Sitter space-time provide other
examples. However, there are many other examples as well; the space of
solutions admitting isolated horizons is in fact \textit{infinite}
dimensional \cite{jl,abf2}. 

All conditions in the definition are \textit{local} to $\Delta$ whence
the isolated horizon can be located quasi-locally; unlike the event
horizon, one does not have to know the \textit{entire} space-time to
determine whether or not a given null surface is an isolated horizon.
Also, there may be gravitational or other radiation arbitrarily close
to $\Delta$. 
\footnote{During this conference Piotr Chruciel pointed out that the
Robinson-Trautman solutions provide examples of exact solutions
which admit non-rotating isolated horizons, have no Killing fields 
and admit radiation arbitrarily close to the horizon.}%
Therefore, in general, space-times admitting isolated
horizons need not be stationary \textit{even in a neighborhood of}
$\Delta$; isolated horizons need not be Killing horizons \cite{jl}.
In spite of this generality, the intrinsic geometry, several of the
curvature components and several components of the Maxwell field at
any isolated horizon are the same as those at the event horizon of
Reissner-Nordstr\"om space-times \cite{abf2,ack,ac}. This similarity
greatly simplifies the detailed analysis.

Finally, isolated horizons are special cases of Hayward's trapping
horizons \cite{sh}, the most important restriction being that the
direction field $[\ell^a]$ is assumed to be
expansion-free. Physically, as explained above, this restriction
captures the idea that the horizon is `isolated', i.e., we are dealing
with an equilibrium situation. The restriction also gives rise to some
mathematical simplifications which, in turn, make it possible to
introduce a well-defined action principle and Hamiltonian framework.
As we will see below, these structures play an essential role in the
proof of the generalized first law and in passage to quantization.

\subsection{Mechanics}
\label{s3.2}

Let me begin by placing the present work on mechanics of isolated
horizons in the context of other treatments in the literature.  The
first treatments of the zeroth and first laws were given by Bardeen,
Carter and Hawking \cite{2} for black holes surrounded by rings of
perfect fluid and this treatment was subsequently generalized to
include other matter sources \cite{mh}.  In all these works, one
restricted oneself to globally stationary space-times admitting event
horizons and considered transitions from one such space-time to a
nearby one. Another approach, based on Noether charges, was introduced
by Wald and collaborators \cite{rw,rwbook}. Here, one again considers
stationary event horizons but allows the variations to be
arbitrary. Furthermore, this method is applicable not only for general
relativity but for stationary black holes in a large class of
theories. In both approaches, the surface gravity $\kappa$ and the
mass $M$ of the hole were defined using the global Killing field and
referred to structure at infinity.

The zeroth and first laws were generalized to arbitrary, non-rotating
isolated horizons $\Delta$ in the Einstein-Maxwell theory in
\cite{abf1,abf2} and dilatonic couplings were incorporated in
\cite{ac}.  In this work, the surface gravity $\kappa$ and the mass
$M_\Delta$ of the isolated horizon refer only to structures
\textit{local} to $\Delta$.%
\footnote{In standard treatments, static solutions are parametrized by
the ADM mass $M$, electric and magnetic charges $Q$ and $P$, dilatonic
charge $D$, cosmological constant $\Lambda$ and the dilatonic coupling
parameter $\alpha$. Of these, $M$ and $D$ are defined \textit{at
infinity}. In the generalized context of isolated horizons, on the
other hand, one must use parameters that are intrinsic to $\Delta$.
Apriori, it is not obvious that this can be done. It turns out that we
can trade $M$ with the area $a_\Delta$ of the horizon and $D$ with the
value $\phi_\Delta$ of the dilaton field on $\Delta$. Boundary
conditions ensure that $\phi_\Delta$ is a constant.}
As mentioned in section \ref{s3.1}, the space ${\cal IH}$ of solutions
admitting isolated horizons is infinite dimensional and static
solutions constitute only a finite dimensional sub-space ${\cal S}$ of
${\cal IH}$. Let us restrict ourselves to the non-rotating case for
comparison. Then, in treatments based on the Bardeen-Carter-Hawking
approach, one restricts oneself only to ${\cal S}$ and variations
tangential to ${\cal S}$. In the Wald approach, one again restricts
oneself to points of ${\cal S}$ but the variations need not be
tangential to ${\cal S}$. In the present approach, on the other
hand, the laws hold at \textit{any} point of ${\cal IH}$ and
\textit{any} tangent vector at that point. However, so far, our results
pertain only to \textit{non-rotating} horizons in a restricted class of
theories.

The key ideas in the present work can be summarized as follows. It is
clear from the setup that surface gravity should be related to the
acceleration of $[\ell^a]$. Recall, however, the acceleration is not a
property of a direction field but of a vector field.  Therefore, to
define surface gravity, we must pick out a specific vector field
$\ell^a$ from the equivalence class $[\ell^a]$. Now, the shear, the
twist, and the expansion of the direction field $[\ell^a]$ all vanish
for \textit{any} choice of normalization. Therefore, we can not use
these fields to pick out a preferred $\ell^a$. However, it turns out
that the expansion $\Theta_{(n)}$ of $n^a$ \textit{is} sensitive to
its normalization. Furthermore, in static solutions, $\Theta_{(n)}$ is
determined entirely by the intrinsic parameters of the
horizon. Therefore, it is natural to require that $\Theta_{(n)}$ be
the same function of the parameters on \textit{any} isolated
horizon. Although it is not apriori obvious, the available rescaling
freedom in the choice of $n$ in fact suffices to meet this requirement
on \textit{any} isolated horizon. Furthermore, the condition
\textit{uniquely} picks out a vector field $n^a$ from the equivalence
class $[n^a]$.  Having a preferred $n^a$ at our disposal, using the
standard normalization $\ell\cdot n = -1$ we can then select an
$\ell^a$ from the equivalence class $[\ell^a]$ uniquely. Finally, we
define surface gravity $\kappa$ to be the acceleration of this
`properly normalized' $\ell^a$; i.e., we set $\ell^a\nabla_a \ell^b =
\kappa \ell^b$ On $\Delta$.

By construction, $\kappa$, so defined, yields the `correct' surface
gravity in the six parameter family of static, dilatonic black-holes.
However, the key question is: Do the zeroth and first laws hold for
general isolated horizons? This is a key test of our strategy of
defining $\kappa$ in the general case. The answer is in the
affirmative.

The zeroth law --constancy of $\kappa$ on isolated horizons-- is
established as follows. First, our boundary conditions on $[\ell^a]$
and $[n^a]$ directly imply that $\kappa$ is constant on each trapped
2-surface. Next, one can show that $\kappa$ can be expressed in terms
of the Weyl curvature component $\Psi_2$ and the expansion
$\Theta_{(n)}$. Finally, the Bianchi identity $\nabla_{[a}R_{bc]de}
=0$, the form of the Ricci tensor component $\Phi_{11}$ dictated by
our boundary conditions on the matter stress-energy, and our
`normalization condition' on $\Theta_{(n)}$ imply that $\kappa$ is
also constant along the integral curves of $\ell^a$. Hence $\kappa$ is
constant on any isolated horizon. To summarize, even though our
boundary conditions allow for the presence of radiation arbitrarily
close to $\Delta$, they successfully extract enough structure
intrinsic to the horizons of static black holes to ensure the validity
of the zeroth law. Our derivation brings out the fact that the zeroth
law is really local to the horizon: Degrees of freedom of the isolated
horizon `decouple' from excitations present elsewhere in space-time.

To establish the first law, one must first introduce the notion of
mass $M_\Delta$ of the isolated horizon. The idea is to define
$M_\Delta$ using the Hamiltonian framework. For this, one needs a
well-defined action principle. Fortunately, even though the boundary
conditions were designed only to capture the notion of an isolated
horizon in a quasi-local fashion, they turn out to be well-suited for
the variational principle. However, just as one must add a suitable
boundary term at infinity to the Einstein-Hilbert action to make it
differentiable in the asymptotically flat context, we must now add
another boundary term at $\Delta$. Somewhat surprisingly, the new
boundary term turns out to be the well-known Chern-Simons action (for
the self-dual connection). This specific form is not important to
classical considerations. However, it plays a key role in the
quantization procedure. The boundary term at $\Delta$ is different
from that at infinity. Therefore one can not simultaneously absorb
both terms in the bulk integral using Stokes' theorem.  Finally, to
obtain a well-defined variational principle for the Maxwell part of
the action, one needs a partial gauge fixing at $\Delta$. One
can follow a procedure similar to the one given above for fixing the
rescaling freedom in $n^a$ and $\ell^a$. It turns out that, not only
does this strategy make the Maxwell action differentiable, but it also
uniquely fixes the scalar potential $\Phi$ at the horizon.

Having the action at one's disposal, one can pass to the Hamiltonian 
framework.%
\footnote{This passage turns out not to be as straightforward as one
might have imagined because there are subtle differences between the
variational principles that lead to the Lagrangian and Hamiltonian 
equations of motion. See \cite{abf2}.}
Now, it turns out that the symplectic structure has, in addition to
the standard bulk term, a surface term at $\Delta$. The surface term
is inherited from the Chern-Simons term in the action and is therefore
precisely the Chern-Simons symplectic structure with a specific
coefficient (i.e., in the language of the Chern-Simons theory, a
specific value of the `level' $k$). The presence of a surface term in
the symplectic structure is somewhat unusual; for example, the
boundary term at infinity in the action does \textit{not} induce a
boundary term in the symplectic structure. 

The Hamiltonian consists of a bulk integral and two surface integrals,
one at infinity and one at $\Delta$. The presence of two surface
integrals is not surprising; for example one encounters it even in the
absence of an internal boundary, if the space-times under
consideration have two asymptotic regions. As usual, the bulk term is
a linear combination of constraints and the boundary term at infinity
is the ADM energy. Using several examples as motivation, we interpret
the surface integral at the horizon as the horizon mass $M_\Delta$
\cite{abf2}. This interpretation is supported by the following result:
If the isolated horizon extends to future time-like infinity $i^+$,
under suitable assumptions one can show that $M_\Delta$ is equal to
the future limit, along $\scri^+$, of the Bondi mass. Finally, note
that $M_\Delta$ is \textit{not} a fundamental, independent attribute
of the isolated horizon; it is a function of the area $a_\Delta$ and
charges $Q_\Delta$, $P_\Delta$, $\phi_\Delta$ which are regarded as
the fundamental parameters.

Thus, we can now assign to any isolated horizon, an area $a_\Delta$, a
surface gravity $\kappa$, an electric potential $\Phi$ and a mass
$M_\Delta$.  The electric charge $Q_\Delta$ can be defined using the
electro-magnetic and dilatonic fields field \textit{at} $\Delta$
\cite{ac}. All quantities are defined in terms of the local structure
at $\Delta$. Therefore, one can now ask: if one moves from
\textit{any} space-time in ${\cal IH}$ to \textit{any} nearby
space-time through a variation $\delta$, how do these quantities vary?
An explicit calculation shows:
$$ \delta M_\Delta = \frac{1}{8\pi G}\,\,\kappa \delta a_\Delta
+ \Phi \delta Q_\Delta\, . $$
(For simplicity, I have restricted myself here to the Einstein-Maxwell
case without dilaton.)  Thus, the first law of black hole mechanics
naturally generalizes to isolated horizons. (As usual, the magnetic
charge can be incorporated via the standard duality rotation.) This
result provides additional support for our strategy of defining
$\kappa$, $\Phi$ and $M_\Delta$.

In static space-times, the mass $M_\Delta$ of the isolated horizon
coincides with the ADM mass $M$ defined at infinity. In general,
$M_\Delta$ is the difference between $M$ and the `radiative energy' of
space-time. However, as in the static case, $M_\Delta$ continues to
include the energy in the `Coulombic' fields ---i.e., the `hair'---
associated with the charges of the horizon, even though it is defined
locally at $\Delta$.  This is a subtle property but absolutely
essential if the first law is to hold in the form given above. To my
knowledge, none of the quasi-local definitions of mass shares this
property with $M_\Delta$. Finally, isolated horizons provide an
appropriate framework for discussing the `physical process version' of
the first law for processes in which the charge of the black hole
changes. The standard strategy of using the ADM mass in place of
$M_\Delta$ appears to run in to difficulties \cite{abf2} and, as far
as I am aware, this issue was never discussed in the literature in the
usual context of context of static event horizons.

\subsection{Quantum geometry in the bulk}
\label{s3.3}

In this sub-section, I will make a detour to introduce the basic ideas
we need from quantum geometry. For simplicity, I will 
ignore the presence of boundaries and focus just on the
structure in the bulk.

There is a common expectation that the continuum picture of space-time,
used in macroscopic physics, would break down at the Planck scale.
This expectation has been shown to be correct within a
non-perturbative, background independent approach to quantum gravity
(see \cite{aa} and references therein).%
\footnote{The necessity of a non-perturbative approach is illustrated
by the following simple example. The energy levels of a harmonic
oscillator are discrete. However, it would be difficult to see this
fundamental discreteness if one were to solve the problem
perturbatively, starting from the Hamiltonian of a free
particle. Similarly, if one \textit{begins} with a continuum
background geometry and then tries to incorporate the quantum effects
perturbatively, it would be difficult to unravel discreteness in the
spectra of geometric operators such as areas of surfaces or volumes of
regions.}
The approach is background independent in the sense that, at the
fundamental level, there is neither a classical metric nor any other
field to perturb around. One only has a bare manifold and \textit{all}
fields, whether they represent geometry or matter, are quantum
mechanical from the beginning.  Because of the subject matter now
under consideration, I will focus on geometry.

Quantum mechanics of geometry has been developed systematically over
the last three years and further exploration continues \cite{aa}.  The
emerging theory is expected to play the same role in quantum gravity
that differential geometry plays in classical gravity.  That is,
quantum geometry is not tied to a specific gravitational theory.
Rather, it provides a kinematic framework or a language to
formulate dynamics in a large class of theories, including general
relativity and supergravity.  In this framework, the fundamental
excitations of gravity/geometry are one-dimensional, rather like 
`polymers' and the continuum picture arises only as an approximation
involving coarse-graining on semi-classical states.  The one
dimensional excitations can be thought of as flux lines of area
\cite{al2}.  Roughly, each line assigns to a surface element it
crosses one Planck unit of area.  More precisely, the area assigned to
a surface is obtained by algebraic operations (involving
group-representation theory) at points where the flux lines intersect
the surface.  As is usual in quantum mechanics, quantum
states of geometry are represented by elements of a Hilbert space
\cite{al1}. I will denote it by $\Hb$.  The basic object for spatial
Riemannian geometry continues to be the triad, but now represented by
an operator(-valued distribution) on $\Hb$ \cite{al2}. All other
geometric quantities ---such as areas of surfaces and volumes of
regions--- are constructed from the triad and represented by
self-adjoint operators on $\Hb$.  The eigenvalues of all geometric
operators are discrete; geometry is thus quantized in the same sense
that the energy and angular momentum of the hydrogen atom are
quantized \cite{al2}.

There is however, one subtlety: there is a one-parameter ambiguity in
this non-perturbative quantization \cite{ip}.  The parameter is
positive, labeled $\gamma$ and called the Immirzi parameter.  This
ambiguity is similar to the $\theta$ ambiguity in the quantization of
Yang-Mills theories.  For all values of $\gamma$, one obtains the same
classical theory, expressed in different canonical variables.
However, quantization leads to a one-parameter family of
\textit{inequivalent} representations of the basic operator algebra.
In particular, in the sector labeled by $\gamma$ the spectra of the
triad ---and hence, all geometric--- operators depend on $\gamma$
through an overall multiplicative factor.  Therefore, while the
qualitative features of quantum geometry are the same in \textit{all}
$\gamma$ sectors, the precise eigenvalues of geometric operators vary
from one sector to another.  The $\gamma$-dependence itself is simple
---effectively, Newton's constant $G$ is replaced by $\gamma G$ in the
$\gamma$-sector.  Nonetheless, to obtain unique predictions, it must
be eliminated and this requires an additional input.  Note however
that since the ambiguity involves a single parameter, as with the
$\theta$ ambiguity in QCD, one judiciously chosen experiment would
suffice to eliminate it.  Thus, for example, if we could measure the
quantum of area , i.e., smallest non-zero value that area of any
surface can have, we would know which value of $\gamma$ is realized in
Nature.  Any further experiment would then be a test of the theory.
Of course, it is not obvious how to devise a feasible experiment to
measure the area quantum directly.  However, we will see that it is
possible to use black hole thermodynamics to introduce suitable
thought experiments.  One of them can determine the value of $\gamma$
and the other can then serve as consistency checks.

\subsection{Quantum geometry of horizon and entropy}
\label{s3.4}

Ideas introduced in the last three sub-sections were combined and
further developed to systematically analyze the quantum geometry of
isolated horizons and calculate their statistical mechanical entropy
in \cite{ack,abck,abk}. (For earlier work, see \cite{kk2,cr}.) In this
discussion, one is interested in space-times with an isolated horizon
with \textit{fixed} values $a_o, Q_o$ and $\phi_o$ of the intrinsic
horizon parameters, the area, the electric charge, and the value of
the dilaton field.

The presence of an isolated horizon $\Delta$ manifests itself in the
classical theory through boundary conditions.  As usual, we can use
some of the boundary conditions to eliminate certain gauge degrees of
freedom at $\Delta$.  The remaining degrees of freedom are coded in an
Abelian connection $V$ defined intrinsically on $\Delta$. $V$ is
constructed from the self-dual spin connection in the bulk.  It is
interesting to note that there are \textit{no} surface degrees of
freedom associated with matter: Given the intrinsic parameters of the
horizon, boundary conditions imply that matter fields defined
intrinsically on $\Delta$ can be completely expressed in terms of
geometrical (i.e., gravitational) fields at $\Delta$.  One can also
see this feature in the symplectic structure.  While the gravitational
symplectic structure acquires a surface term at $\Delta$, matter
symplectic structures do not. We will see that this feature provides a
simple explanation of the fact that, among the set of intrinsic
parameters natural to isolated horizons, entropy depends only on area.

Of particular interest to the present Hamiltonian approach is the
pull-back of $V$ to the 2-sphere $S_\Delta$ (orthogonal to $\ell^a$
and $n^a$) at which the space-like 3-surfaces $M$ used in the phase
space construction intersect $\Delta$. (See figure 1(a).) This
pull-back ---which I will also denote by $V$ for simplicity--- is
precisely the $U(1)$ spin-connection of the 2-sphere $S_\Delta$.  Not
surprisingly, the Chern-Simons symplectic structure for the
non-Abelian self-dual connection that I referred to in Section
\ref{s3.2} can be re-expressed in terms of $V$.  The result is
unexpectedly simple \cite{ack}: the surface term in the total symplectic
structure is now just the Chern-Simons symplectic structure for the
\textit{Abelian} connection $V$!  The only remaining boundary
condition relates the curvature $F= dV$ of $V$ to the triad vectors.
This condition is taken over as an operator equation.  Thus, in the
quantum theory, neither the intrinsic geometry nor the curvature of
the horizon are frozen; neither is a classical field.  Each is allowed
to undergo quantum fluctuations but because of the operator equation
relating them, they have to fluctuate in tandem.

To obtain the quantum description in presence of isolated horizons, 
therefore, one begins with a fiducial Hilbert space $\H = \Hb\otimes\Hs$ 
where $\Hb$ is the Hilbert space associated with the bulk polymer geometry 
and $\Hs$ is the Chern-Simons Hilbert space for the connection $V$.%
\footnote{In the classical theory, all fields are smooth, whence the value
of any field in the bulk determines its value on $\Delta$ by continuity.
In quantum theory, by contrast, the measure is concentrated on generalized
fields which can be arbitrarily discontinuous, whence surface states 
are no longer determined by bulk states. A compatibility relation 
does exist but it is introduced by the quantum boundary condition. It 
ensures that the total state is invariant under the permissible internal 
rotations of triads.}
The quantum boundary condition says that only those states in $\H$ are
allowed for which there is a precise intertwining between the bulk and
the surface parts.  However, because the required intertwining is
`rigid', apriori it is not clear that the quantum boundary conditions
would admit \textit{any} solutions at all.  For solutions to exist,
there has to be a very delicate matching between certain quantities on
$\Hb$ calculated from the bulk quantum geometry and certain quantities
on $\Hs$ calculated from the Chern-Simons theory.  The precise
numerical coefficients in the surface calculation depend on the
numerical factor in front of the surface term in the symplectic
structure (i.e., on the Chern-Simons level $k$) which is itself
determined in the classical theory by the coefficient in front of the
Einstein-Hilbert action and our classical boundary conditions.  Thus,
the existence of a coherent quantum theory of isolated horizons
requires that the three corner stones ---classical general relativity,
quantum mechanics of geometry and Chern-Simons theory--- be united
harmoniously.  Not only should the three conceptual frameworks fit
together seamlessly but certain \textit{numerical coefficients},
calculated independently within each framework, have to match
delicately.  Fortunately, these delicate constraints are met and the
quantum boundary conditions admit a sufficient number of solutions.

Because we have fixed the intrinsic horizon parameters, is is natural
to construct a micro-canonical ensemble from eigenstates of the
corresponding operators with eigenvalues in the range $(q_o -\delta q
, q_o +\delta q)$ where $\delta q$ is very small compared to the fixed
value $q_o$ of the intrinsic parameters. Since there are no surface
degrees of freedom associated with matter fields, let us focus on
area, the only gravitational parameter available to us. Then, we only
have to consider those states in $\Hb$ whose polymer excitations
intersect $S_\Delta$ in such a way that they endow it with an area in
the range $(a_o -\delta a , a_o + \delta a)$ where $\delta a$ is of
the order of $\l^2$ (with $\l$, the Planck length).  Denote by $\P$
the set of punctures that any one of these polymer states makes on
$S_\Delta$, each puncture being labeled by the eigenvalue of the area
operator at that puncture.  Given such a bulk state, the quantum
boundary condition tells us that only those Chern-Simons surface
states are allowed for which the curvature is concentrated at
punctures and the range of allowed value of the curvature at each
puncture is dictated by the area eigenvalue at that puncture. Thus,
for each $\P$, the quantum boundary condition picks out a sub-space
$\Hs^{\P}$ of the surface Hilbert space $\Hs$. Thus, the quantum
geometry of the isolated horizon is effectively described by states in
$$\Hs^{\rm phys} = \bigoplus_{\P} \Hs^{\P} $$
as $\P$ runs over all possible punctures and area-labels at each 
puncture, compatible with the requirement that the total area assigned 
to $S_\Delta$ lie in the given range.

\begin{figure}
  \begin{center}
    \begin{minipage}{3in}
      \begin{center}
        \psfrag{ji}{$j_i$}
    	\psfrag{mi}{$m_i$}
	\psfrag{S}{$S_\Delta$}
        \includegraphics[height=6cm]{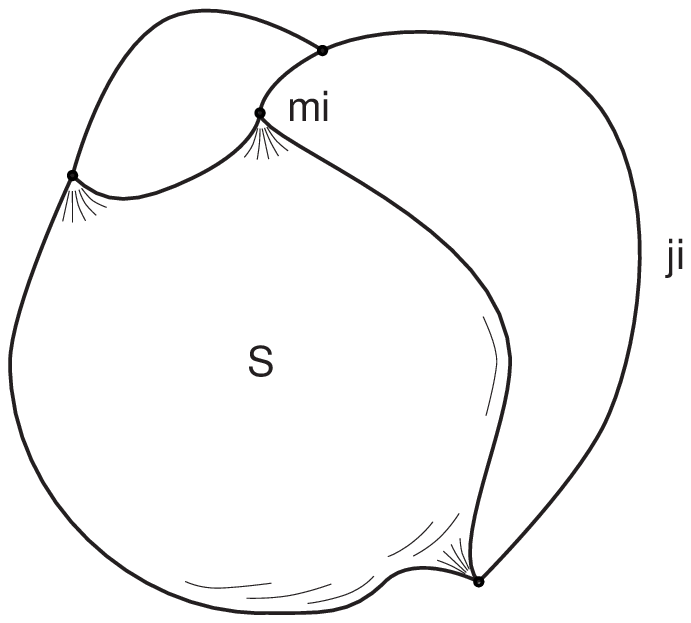}\\(a)
      \end{center}
    \end{minipage}
    \hspace{.5in}
    \begin{minipage}{2.5in}
      \begin{center}
	\psfrag{pi}{$p_i$}
	\psfrag{gamma}{$\gamma$}
        \includegraphics[width=4cm]{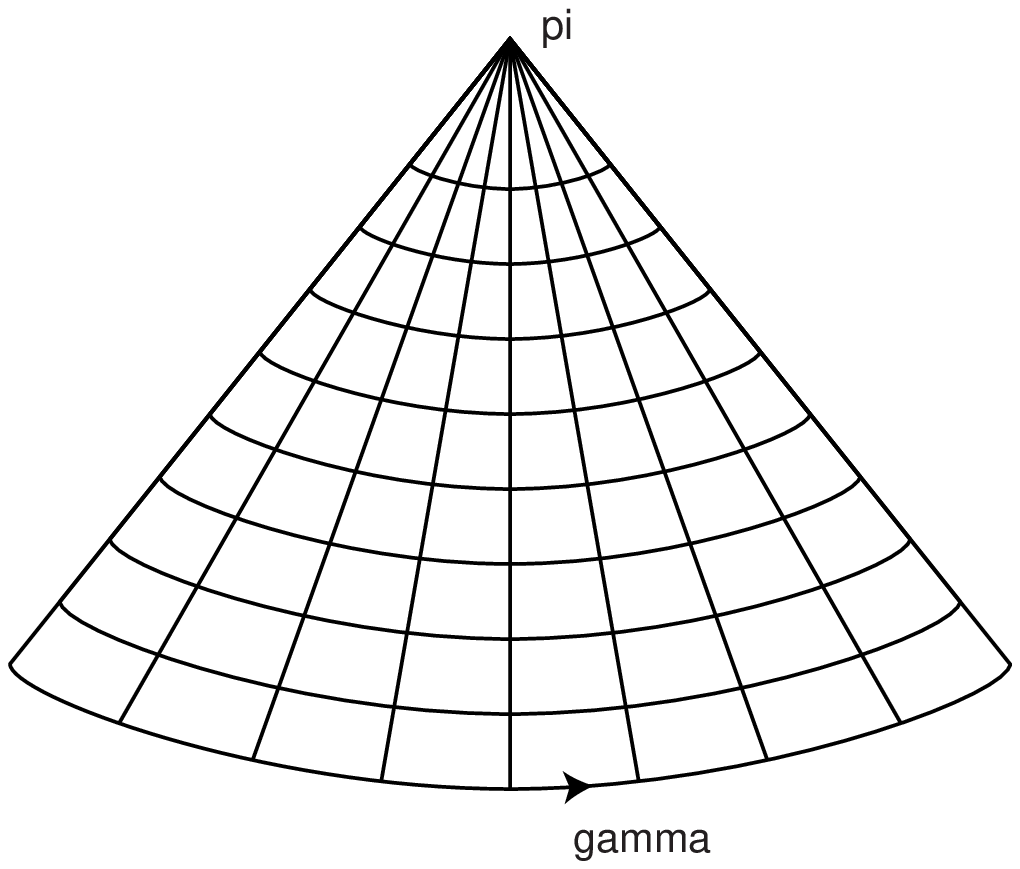}\\(b)
      \end{center}
    \end{minipage}
\caption{(a)\quad Quantum geometry around an isolated horizon. The
$i$-th polymer excitation of the bulk geometry carries a $1/2$-integer
label $j_i$. Upon puncturing the horizon 2-sphere $S_\Delta$, it
induces $8\pi \gamma \sqrt{j_i(j_i+1)}$ Planck units of area.  At each
puncture, in the intrinsic geometry of $S_\Delta$, there is a deficit
angle of $2\pi m_i/k$, where $m_i$ is a $1/2$-integer in the interval
$[-j_i, j_i]$ and $k$ the `level' of the Chern-Simons theory.  \quad
(b)\quad Magnified view of a puncture $p_i$.  The holonomy of the
$U(1)$ connection $V$ around a loop $\gamma$ surrounding any puncture
$p_i$ determines the deficit angle at $p_i$. Each deficit angle is
quantized and they add up to $2\pi$.}
\label{fig3}
\end{center}
\end{figure}

One can visualize this quantum geometry as follows.  Given any one
state in $\Hs^{\P}$, the connections $V$ are flat everywhere except at
the punctures and the holonomy around each puncture is fixed.  Using
the classical interpretation of $V$ as the metric compatible spin
connection on $S_\Delta$ we conclude that, in quantum theory, the
intrinsic geometry of the horizon is flat except at the punctures.  At
each puncture, there is a deficit angle, whose value is determined by
the holonomy of $V$ around that puncture.  Since each puncture
corresponds to a polymer excitation in the bulk, polymer lines can be
thought of as `pulling' on the horizon, thereby producing deficit
angles in an otherwise flat geometry (see figure \ref{fig3}).  Each
deficit angle is quantized and the angles add up to $2\pi$ as in a
discretized model of a 2-sphere geometry. Thus, the quantum geometry
of an isolated horizon is quite different from its smooth classical
geometry.  In addition, of course, each polymer line endows the
horizon with a small amount of area and these area elements add up to
provide the horizon with total area in the range $(a_0 -\delta a,
a_o+\delta a)$.  Thus, one can intuitively picture the quantum horizon
as the surface of a large, water-filled balloon which is suspended
with a very large number of wires, each exerting a small tug on the
surface at the point of contact and giving rise to a `conical
singularity' in the geometry.

Finally, one can calculate the entropy of the quantum micro-canonical
ensemble. We are not interested in the \textit{full} Hilbert space
since the `bulk-part' includes, e.g., states of gravitational
radiation and matter fields far away from $\Delta$. Rather, we wish to
consider only the states of the isolated horizon $\Delta$
itself. Therefore, we are led to trace over the `bulk states' to
construct a density matrix $\rho_{\rm IH}$ describing a
maximum-entropy mixture of surface states for which the intrinsic
parameters lie in the given range. The statistical mechanical entropy
is then given by $ S= -{\rm Tr} \, \rho_{\rm IH} \ln \rho_{\rm
IH}$. As usual, the trace can be obtained simply by counting states,
i.e., by computing the dimension $\N$ of $\Hs^{\rm phys}$. We have:
$$\N = \exp\, (\frac{\gamma_o}{\gamma}\,\frac{a_o}{4\l^2})
\quad\quad {\rm where} \quad\quad \gamma_o = \frac{\ln 2}{\pi\sqrt{3}}
$$
Thus, the number of micro-states does go exponentially as area. This
is a non-trivial result. For example if, as in the early treatments,
one ignores boundary conditions and the Chern-Simons term in the
symplectic structure and does a simple minded counting, one finds that
the exponent in $\N$ is proportional to $\sqrt{a_o}$.  However, our
numerical coefficient in front of the exponent depends on the Immirzi
parameter $\gamma$. The appearance of $\gamma$ can be traced back
directly to the fact that, in the $\gamma$-sector of the theory, the
area eigenvalues are proportional to $\gamma$.  Thus, because of the
quantization ambiguity, the $\gamma$-dependence of $\N$ is inevitable.

We can now adopt the following `phenomenological' viewpoint.  In the
infinite dimensional space ${\cal IH}$, one can fix one space-time
admitting isolated horizon, say the Schwarzschild space-time with mass
$M_o >> M_{\rm Pl}$, (or, the de Sitter space-time with the
cosmological constant $\Lambda_o << 1/\l^2$).  For agreement with
semi-classical considerations, in these cases, entropy should be given
by $S = ({a_o}/{4\l^2})$ which can happen only in the sector $\gamma =
\gamma_o$ of the theory. The theory is now completely determined
and we can go ahead and calculate the entropy of any other isolated
horizon in \textit{this} theory. Clearly, we obtain:
$$S_{\rm IH} = \frac{1}{4}\, \frac{a_o}{\l^2} $$
for \textit{all} isolated horizons. Furthermore, in this
$\gamma$-sector, the statistical mechanical temperature of any
isolated horizon is given by Hawking's semi-classical value
$\kappa\hbar/2\pi$ \cite{ak,kk2}. Thus, we can do one thought
experiment ---observe the temperature of a large black black hole from
far away--- to eliminate the Immirzi ambiguity and fix the theory.
This theory then predicts the correct entropy and temperature for all
isolated horizons in ${\cal IH}$ with $a_o >> \l^2$.

The technical reason behind this universality is trivial.  However, the
conceptual argument is not because it is quite non-trivial that $\N$
depends only on the area and not on values of other charges.
Furthermore, the space ${\cal IH}$ is infinite dimensional and it is
not apriori obvious that one should be able to give a statistical
mechanical account of entropy of \textit{all} isolated horizons in one
go.  Indeed, values of fields such as $\Psi_4$ and $\phi_2$ can be
vary from one isolated horizon to another even when they have same
intrinsic parameters. This freedom could well have introduced
obstructions, making quantization and entropy calculation impossible.
That this does not happen is related to but independent of the fact
that this feature did not prevent us from extending the laws of
mechanics from static event horizons to general isolated horizons.

I will conclude this sub-section with two remarks.

i) In this approach, we began with the sector of general relativity
admitting isolated horizons and then quantized that sector. Therefore,
ours is an `effective' description. In a fundamental description, one
would begin with the full quantum theory and isolate in it the sector
corresponding to quantum horizons. Since the notion of horizon is
deeply tied to classical geometry, at the present stage of our
understanding, this goal appears to be out of reach in all approaches
to quantum gravity. However, for thermodynamic considerations of large
horizons, the effective description should be sufficient.

ii) The notion of entropy used here has two important features.
First, in this framework, the notion is not an abstract property of
the space-time as a whole but depends on the division of space-time in
to an exterior and an interior. Operationally, it is tied to the class
of observers who live in the exterior region for whom the isolated
horizon is a {\it physical} boundary that separates the part of the
space-time they can access from the part they can not.  (This is in
sharp contrast to early work which focussed on the interior.)  This
point is especially transparent in the case of cosmological horizons
in de Sitter space-time since that space-time does not admit an
invariantly defined division. The second feature is that, although
there is `observer dependence' in this sense, the entropy does {\it
not} refer to the degrees of freedom in the interior. Indeed, nowhere
in our calculation did we analyze the states associated with the
interior.  Rather, our entropy refers to the micro-states of the
boundary itself which are compatible with the macroscopic constraints
on the area and charges of the horizon; it counts the physical
micro-states which can interact with the outside world, not
disconnected from it.

\section{Discussion}

Perhaps the most pleasing aspect of this analysis is the existence of
a single framework to encompass diverse ideas at the interface of
general relativity, quantum theory and statistical mechanics.  In the
classical domain, this framework generalizes laws of black hole
mechanics to physically more realistic situations.  At the quantum
level, it provides a detailed description of the quantum geometry of
horizons and leads to a statistical mechanical calculation of entropy.
In both domains, the notion of isolated horizons provides an unifying
arena enabling us to handle different types of situations ---e.g.,
black holes and cosmological horizons--- in a single stroke.  In the
classical theory, the same line of reasoning allows one to establish
the zeroth and first laws for \textit{all} isolated horizons.
Similarly, in the quantum theory, a single procedure leads one to
quantum geometry and entropy of \textit{all} isolated horizons.  By
contrast, in other approaches, fully quantum mechanical treatments
seem to be available only for stationary black holes.  Indeed, to my
knowledge, even in the static case, a complete statistical mechanical
calculation of the entropy of cosmological horizons has not been
available.  Finally, our extension of the standard \textit{Killing}
horizon framework sheds new light on a number of issues, particularly
the notion of mass of associated to an horizon and the physical
process version of the first law \cite{abf2}.

However, the framework presented here is far from being complete and
provides promising avenues for future work.  First, while some of the
motivation behind our approach is similar to the considerations that
led to the interesting series of papers by Brown and York \cite{by},
not much is known about the relation between the two frameworks. It
would be interesting to explore this relation, and more generally, to
relate the isolated horizon framework to the semi-classical ideas
based on Euclidean gravity.  Second, while the understanding of the
micro-states of an isolated horizon is fairly deep by now, work on a
quantum gravity derivation of the Hawking radiation is still in a
preliminary stage.  Using general arguments based on Einstein's A and
B coefficients \cite{1} and the known micro-states of an isolated
horizon, one can argue \cite{kk1} that the envelope of the line
spectrum emitted by a black hole should be thermal.  However, further
work is necessary to make sure that the details are correct.

For the laws of mechanics and the entropy calculation, the obvious
open problem is the extension to incorporate non-zero angular
momentum. Recently, Jerzy Lewandowski has performed an exhaustive
analysis of the geometrical structure of general isolated horizons and
streamlined the necessary background material. further recent work in
collaboration with him and with Chris Beetle and Steve Fairhurst has
led to a generalization of boundary conditions to incorporate rotation
(as well as distortion in absence of rotation) and a proof of the
zeroth law in the general context. Construction of the corresponding
Hamiltonian framework is now under way.  The extension of the entropy
calculation, on the other hand, may turn out to be trickier for it may
well require a new technical insight.  On a long range, the
outstanding challenge is to obtain a deeper understanding of the
Immirzi ambiguity and the associated issue of renormalization of
Newton's constant.  For any value of $\gamma$, one obtains the
`correct' classical limit.  However, as far as black hole
thermodynamics is concerned, it is only for $\gamma=\gamma_o$ that one
seems to obtain agreement with quantum field theory in curved
space-times.  Is this value of $\gamma$ robust?  Can one make further
semi-classical checks?  A pre-requisite for this investigation is a
better handle on the issue of semi-classical states.  A major effort
will soon be devoted to this issue.

Let me conclude with a comparison between the entropy calculation in
this approach and those performed in string theory.  First, there are
some obvious differences.  In the present approach, one begins with
the sector of the classical theory containing space-times with
isolated horizons and then proceeds with quantization. Consequently,
one can keep track of the physical, curved geometry.  In particular,
as required by physical considerations, the micro-states which account
for entropy can interact with the physical exterior of the black hole.
In string theory, by contrast, actual calculations are generally
performed in flat space and non-renormalization arguments and/or
duality conjectures are then invoked to argue that the results so
obtained refer to macroscopic black holes.  Therefore, relation to the
curved space geometry and physical meaning of the degrees of freedom
which account for entropy is rather obscure.  More generally, lack of
direct contact with physical space-time can also lead to practical
difficulties while dealing with other macroscopic situations.  For
example, in string theory, it may be difficult to account for the
entropy normally associated with de Sitter horizons.  On the other
hand, in the study of genuinely quantum, Planck size black holes, this
`distance' from the curved space-time geometry may turn out to be a
blessing, as classical curved geometry will not be an appropriate tool
to discuss physics in these situations.  In particular, a description
which is far removed from space-time pictures may be better suited in
the discussion of the last stages of Hawking evaporation and the
associated issue of `information loss'.

The calculations based on string theory have been carried out in a
number of space-time dimensions while the approach presented here is
directly applicable only to four dimensions. An extension of the
underlying non-perturbative framework to higher dimensions was
recently proposed by Freidel, Krasnov and Puzzio but a systematic
development of quantum geometry has not yet been undertaken. Also, our
quantization procedure has an inherent $\gamma$-ambiguity which
trickles down to the entropy calculation.  By contrast, calculations
in string theory are free of this problem.  On the other hand, most
detailed calculations in string theory have been carried out only for
(a sub-class of) extremal or near-extremal black holes.  While these
black holes are especially simple to deal with mathematically,
unfortunately, they are not of direct relevance to astrophysics, i.e.,
to the physical world we live in.  More recently, using the Maldecena
conjecture, stringy calculations have been extended to non-extremal
black holes with $R^2_{\rm Sch} >> 1/\Lambda$, where $R_{\rm Sch}$ is
the Schwarzschild radius. However, the numerical coefficient in front
of the entropy turns out to be incorrect and it is not yet clear
whether inclusion of non-Abelian interactions, which are ignored in
the current calculations, would restore the numerical coefficient to
its correct value.  Furthermore, it appears that a qualitatively new
strategy may be needed to go beyond the $R^2_{\rm Sch} >> 1/\Lambda$
approximation.  Finally, as in other results based on the Maldecena
conjecture, the underlying boundary conditions at infinity are quite
unphysical since the radius of the compactified dimensions is required
to equal the cosmological radius even near infinity. Hence the
relevance of these mathematically striking results to our physical
world remains unclear.  In the current approach, by contrast,
ordinary, astrophysical black holes in the physical, four space-time
dimensions are included from the beginning.

In spite of this differences, there are some striking similarities.
Our polymer excitations resemble stings.  Our horizon looks like a
`gravitational 2-brane'.  Our polymer excitations ending on the
horizon, depicted in figure \ref{fig3}, closely resemble strings with
end points on a membrane.  As in string theory, our `2-brane' carries
a natural gauge field.  Furthermore, the horizon degrees of freedom
arise from this gauge field.  These similarities seem astonishing.
However, a closer look brings out a number of differences as well.  In
particular, being horizon, our `2-brane' has a direct interpretation
in terms of the curved \textit{space-time geometry} and our $U(1)$
connection is the \textit{gravitational} spin-connection on the
horizon.  Nonetheless, it may well be that, when quantum gravity is
understood at a deeper level, it will reveal that the striking
similarities are not accidental, i.e., that the two descriptions are in
fact closely related.

\bigskip\bigskip

\noindent\textbf{Acknowledgments:} The material presented in this
report is based on joint work with John Baez, Chris Beetle, Alex
Corichi, Steve Fairhurst and especially Kirill Krasnov.  I am grateful
to them for collaboration and countless discussions. Special thanks
are due to Chris Beetle for his help with figures. I would like to
thank Jerzy Lewandowski for sharing his numerous insights on the
isolated horizon boundary conditions.  I have also profited from
comments made by Brandon Carter, Piotr Chrusciel, Helmut Friedrich,
Sean Hayward, Gary Horowitz, Ted Jacobson, Don Marolf, Jorge Pullin,
Istavan Racz, Oscar Reula, Carlo Rovelli, Bernd Schmidt, Daniel
Sudarsky, Thomas Thiemann and Robert Wald.  This work was supported in
part by the NSF grants PHY94-07194, PHY95-14240, INT97-22514 and by
the Eberly research funds of Penn State.

\end{document}